\documentclass[prb,twocolumn,showpacs,preprintnumbers]{revtex4}
\usepackage{bm}
\usepackage{amsmath}
\usepackage[dvips]{graphicx}
\usepackage{color}
\usepackage{longtable}
\usepackage{epsfig}
\usepackage{color}

\begin{document}

\title{Spin-stiffness of anisotropic Heisenberg model on square lattice and
possible mechanism for pinning of the electronic liquid crystal direction in
YBCO}
\author{T. Pardini}
\affiliation{Physics Department, University California Davis, CA 95616, USA}
\author{R. R. P. Singh}
\affiliation{Physics Department, University California Davis, CA 95616, USA}
\author{A. Katanin}
\affiliation{Institute of Metal Physics, Kovalevskaya str. 18, 620041, Ekaterinburg,
Russia}
\author{O. P. Sushkov}
\affiliation{School of Physics, University of New South Wales, Sydney 2052, Australia}

\begin{abstract}
Using series expansions and spin-wave theory we calculate the spin-stiffness 
anisotropy $\rho_{sx}/\rho_{sy}$ in Heisenberg models on the square lattice with anisotropic couplings $J_x,J_y$. 
We find that for the weakly anisotropic spin-half model ($J_x\approx J_y$), $\rho_{sx}/\rho_{sy}$
 deviates substantially from the naive
estimate $\rho_{sx}/\rho_{sy} \approx J_x/J_y$. We argue that this deviation
can be responsible for pinning the electronic liquid crystal direction, a
novel effect recently discovered in YBCO. For completeness, we also study
the spin-stiffness for arbitrary anisotropy $J_x/J_y$ for spin-half and
spin-one models. In the limit of $J_y/J_x\to 0$, when the model reduces to
weakly coupled chains, the two show dramatically different behavior. In the
spin-one model, the stiffness along the chains goes to zero, implying the
onset of Haldane-gap phase, whereas for spin-half the stiffness along the
chains increases monotonically from a value of $0.18 J_x$ for $J_y/J_x=1$
towards $0.25 J_x$ for $J_y/J_x\to 0$. Spin-wave theory is extremely
accurate for spin-one but breaks down for spin-half presumably due to the onset of
topological terms.
\end{abstract}

\maketitle

%\date{\today}

%\pacs{
%To be determined later
%75.10.Jm, 75.30.Ds 
%}
This work is motivated by the recent discovery~\cite{Hinkov2007,Hinkov2008} 
of the
electronic liquid crystal in underdoped cuprate superconductor YBa$_2$Cu$_3$O%
$_{6.45}$. The electronic liquid crystal manifests itself in a strong
anisotropy of the low energy inelastic neutron scattering. The liquid
crystal picture implies a spontaneous violation of the directional symmetry:
the ``crystal'' can be oriented either along the (1,0) or along the (0,1) axes of
the square lattice. The YBa$_2$Cu$_3$O$_{6.45}$ compound has a tetragonal
lattice with tiny in-plane lattice anisotropy, ${a^*}/{b^*} \approx 0.99$.
This tiny anisotropy is sufficient to pin the orientation of the electronic
liquid crystal along the $a^*$-axis.
As a result the low energy  neutron 
scattering~\cite{Hinkov2007,Hinkov2008} 
demonstrates a quasi-1D structure along $a^*$.

To understand the pinning mechanism of the electronic crystal, in the present work, we study 
the anisotropic Heisenberg model. We calculate the in-plane
anisotropy of the spin-stiffness and demonstrate that this is
strongly enhanced by quantum fluctuations. We argue that the enhancement is
sufficient to provide a pinning mechanism for the initially spontaneous
orientation of the electronic liquid crystal and suggest a specific
mechanism for the pinning.

The anisotropic Heisenberg model has previously attracted a lot of
theoretical interest\cite{Affleck1996,Affleck1994,Parola1993,Sandvik1999,Katanin2000}.
However, most theoretical studies have focussed on the regime of strong
anisotropy, where the system reduces to one of weakly coupled spin-chains,
and the most significant issue there is that of dimensional crossover and
the onset of long range antiferromagnetic order. To the best of our
knowledge the anisotropy of spin-stiffness has not been studied before. This
is an important theoretical problem in itself and therefore we extend our
study to the case of arbitrary strong anisotropy. We consider both spin-half 
model, where in the limit of strong anisotropy we come to the situation
of weakly coupled Heisenberg S=1/2 chains, and also spin-one model, where in the
limit of strong anisotropy we come to the situation of weakly coupled Haldane chains.

Our series expansion results show 
that the spin-stiffness indeed behaves very differently in the two cases.
For spin-one, the stiffness
along the chains vanishes at an anisotropy ratio of $J_y/J_x\approx 0.01$.
Self-consistent spin-wave theory remains highly accurate
in this case all the way down to the transition. On the other hand,
for spin-half, series expansions show that the stiffness along the chains
increases from $0.18 J_x$ in the isotropic limit towards the known\cite{Shastry1990}
1D result of $0.25 J_x$ as $J_y\to 0$. In this case
 spin-wave theory clearly breaks down with increasing anisotropy, presumably due to
the onset of Berry phase interference\cite{Sachdev1995}.

The structure of the paper is as follows: in Section~\ref{SE} we calculate the
spin-stiffness using series expansions. This is probably the most accurate
method that is valid from small to very large anisotropy. In Section~\ref{SW} 
we calculate the same spin-stiffness using spin-wave theory.
This method is valid as long as one is not close to 1D limit. In Section~\ref{pinning} we
discuss the application of our results to the explanation of the electronic liquid
crystal pinning in YBa$_2$Cu$_3$O$_{6.45}$. Finally, in Section~\ref{conclusions}, we draw our conclusions.

\section{Hamiltonian and series calculation}\label{SE}
\begin{figure}[ht]
  \label{spinhalf} \resizebox{85mm}{!}{\includegraphics{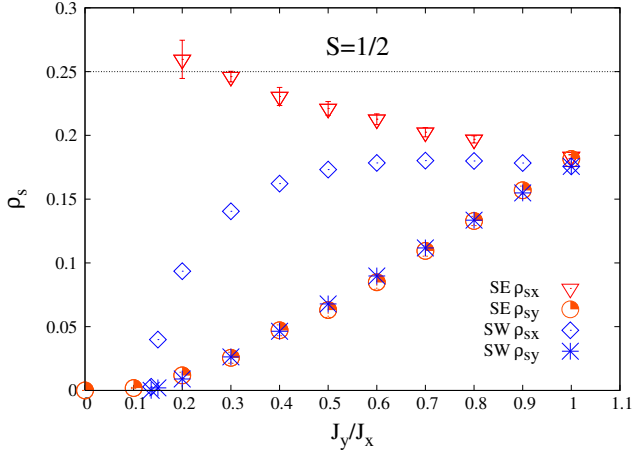}}
  \caption{(Color online) Series Expansion (SE) and Spin-Wave (SW) spin-stiffness of the spin-half Heisenberg model on
    the anisotropic square lattice along the \textit{x} and \textit{y} axis as a
    function of the anisotropy $J_y/J_x$. The dotted line shows the value
    of the spin-stiffness $\protect\rho_{sx}$ in the 1D limit of the model ($J_y/J_x=0$).}
  \label{fig:spinhalf}
\end{figure}
\begin{figure}[ht]
  \label{spinone} \resizebox{85mm}{!}{\includegraphics{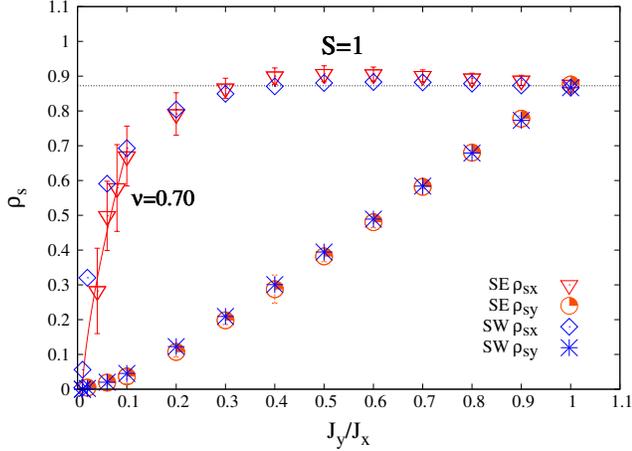}}
\caption{(Color online) Series Expansion (SE) and Spin-Wave (SW) spin-stiffness of the spin-one Heisenberg model on
the anisotropic square lattice along the \textit{x} and \textit{y} axis as a
function of the anisotropy $J_y/J_x$. In the region $0.01\leq J_y\leq
0.1$ the data points for the SE $\protect\rho_{sx}$ have been fitted to the curve $%
\protect\rho_{sx}=(J_y-0.01)^\protect\protect\nu$, with $\protect\nu=0.7\pm
0.1$. The dotted line represents the spin-stiffness in the 2D isotropic
limit of the model ($J_y/J_x=1$).}
\label{fig:spinone}
\end{figure}
\begin{figure}[ht]\label{alpha} 
\resizebox{85mm}{!}{\includegraphics{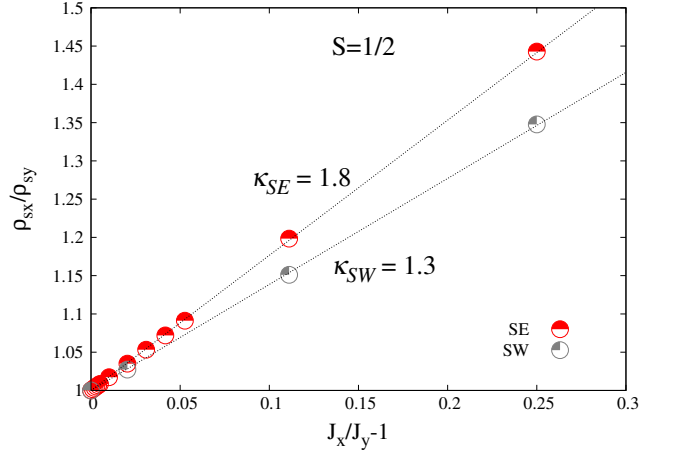}}
\caption{(Color online) Series Expansion (SE) and Spin-Wave (SW) spin-stiffness anisotropy 
  $\protect\rho_{sx}/\protect\rho_{sy}$
  as a function of $J_x/J_y-1$ for the spin-half Heisenberg model on
  the anisotropic square lattice. The data points have been fitted to the linear
  function given in Eq. (\protect\ref{linearfit}) yielding $\protect\kappa_{SE} =1.8$ and 
  $\protect\kappa_{SW} =1.3$}
\label{fig:alpha}
\end{figure}
\squeezetable     
\begin{table}[tbp]
\caption{Series expansion coefficients for the spin-stiffness of the
spin-1/2 Heisenberg model on the anisotropic square lattice for selected
values of the anisotropy coupling $J_y$.}
\label{TABLE 1}%
\begin{ruledtabular}
    \begin{tabular}{ccccc}
      order&$J_x$&$J_y$&$\rho_{sx}$&$\rho_{sy}$\\
      \hline
      0&1&0.9&0.2500000000&0.2250000000\\
      1&&&0.0892857142&0.0698275862\\
      2&&&-0.0927175942&-0.0926628428\\
      3&&&-0.0151861349&-0.0132149279\\
      4&&&-0.0045297105&-0.0011224771\\
      5&&&0.0003357042&0.0021422839\\
      6&&&-0.0057755038&-0.0049661531\\
      7&&&-0.0003145877&-0.0002219010\\
      8&&&-0.0042654841&-0.0038964097\\
      \hline
      0&1&0.7&0.2500000000&0.1750000000\\
      1&&&0.1041666666&0.0453703703\\
      2&&&-0.0868950718&-0.0850367899\\
      3&&&-0.0172969207&-0.0105492432\\
      4&&&-0.0045937341&0.0050836143\\
      5&&&-0.0012261711&0.0039781766\\
      6&&&-0.0059912113&-0.0040000116\\
      7&&&-0.0006353314&-0.0005917251\\
      8&&&-0.0045520522&-0.0033419898\\
      \hline
      0&1&0.5&0.2500000000&0.1250000000\\
      1&&&0.1250000000&0.0250000000\\
      2&&&-0.0891666666&-0.0779166666\\
      3&&&-0.0235044642&-0.0081537698\\
      4&&&0.0013212991&0.0140706091\\
      5&&&-0.0017270061&0.0056040091\\
      6&&&-0.0060489454&-0.0045071391\\
      7&&&-0.0009351484&-0.0017589857\\
      8&&&-0.0057105087&-0.0027490512\\
      \hline
      0&1&0.3&0.2500000000&0.0750000000\\
      1&&&0.1562500000&0.0097826086\\
      2&&&-0.1094346417&-0.0666305597\\
      3&&&-0.0408511573&-0.0052660964\\
      4&&&0.0234241580&0.0242642206\\
      5&&&0.0016805213&0.0056385893\\
      6&&&-0.0085614425&-0.0079166392\\
      7&&&-0.0010470671&-0.0031407541\\
      8&&&-0.0085663482&-0.0003268242\\
    \end{tabular}
  \end{ruledtabular}
\end{table}
\squeezetable     
\begin{table}[tbp]
\caption{Series expansion coefficients for the spin-stiffness of the spin-1
Heisenberg model on the anisotropic square lattice for selected values of
the anisotropy coupling $J_y$.}
\label{TABLE 2}%
\begin{ruledtabular}
    \begin{tabular}{ccccc}
      order&$J_x$&$J_y$&$\rho_{sx}$&$\rho_{sy}$\\
      \hline
      0&1&0.9&1.0000000000&0.9000000000\\
      1&&&0.1515151515&0.1208955223\\
      2&&&-0.1535289080&-0.1422381534\\
      3&&&0.0207519434&0.0193249175\\
      4&&&-0.0428180052&-0.0391942098\\
      5&&&0.0072984862&0.0068989716\\
      6&&&-0.0210286555&-0.0190202422\\
      7&&&0.0043677522&0.0041430202\\
      \hline
      0&1&0.7&1.0000000000&0.7000000000\\
      1&&&0.1724137931&0.0803278688\\
      2&&&-0.1553008729&-0.1192091787\\
      3&&&0.0196401862&0.0152963896\\
      4&&&-0.0421528844&-0.0313929868\\
      5&&&0.0066352559&0.0055840695\\
      6&&&-0.0210149617&-0.0149157761\\
      7&&&0.0040511283&0.0034008244\\
      \hline
      0&1&0.5&1.0000000000&0.5000000000\\
      1&&&0.2000000000&0.0454545454\\
      2&&&-0.1688941361&-0.0982465564\\
      3&&&0.0182177620&0.0106543293\\
      4&&&-0.0421971280&-0.0242928541\\
      5&&&0.0050652115&0.0041087839\\
      6&&&-0.0210666888&-0.0106538747\\
      7&&&0.0035338724&0.0025923026\\
      \hline
      0&1&0.3&1.0000000000&0.3000000000\\
      1&&&0.2380952380&0.0183673469\\
      2&&&-0.2069165600&-0.0746342335\\
      3&&&0.0173382982&0.0055030901\\
      4&&&-0.0479905685&-0.0189077310\\
      5&&&0.0002743242&0.0023851969\\
      6&&&-0.0213911544&-0.0061871738\\
      7&&&0.0021597090&0.0016832884\\
    \end{tabular}
  \end{ruledtabular}
\end{table}
We consider Antiferromagnetic Heisenberg model on a square-lattice, with
spatially anisotropic exchange couplings given by the Hamiltonian 
\begin{equation}  \label{hamiltonian}
\mathcal{H}=J_x\sum_{\vec{r}} {\vec{S}_{\vec {r}}\cdot {\vec {S}_{\vec {r}+%
\hat {x}} +J_y\sum_{\vec {r}} {\vec {S}_{\vec {r}}\cdot {\vec {S}_{\vec {r}+%
\hat {y}}}}}} \ ,
\end{equation}
where the sum over $\vec r$ runs over all sites of the square-lattice.
Spin-stiffness can be defined by the change in the ground state energy of
the system under an applied twist along one of the axes\cite%
{Singh1989,Hamer1994}. In general, it can be decomposed into a sum of two
parts, a paramagnetic part and a diamagnetic part. For the anisotropic
model, one can define two different twists $\rho_{sx}$ and $\rho_{sy}$
depending on whether the twist is applied along the $x$ or the $y$ axis.
Following Ref.~\onlinecite{Singh1989,Hamer1994}, the diamagnetic component of the
twist for $\rho_{sy}$ is given by the expression 
\begin{equation}
\rho_{sy}^{dia}=-J_y<S_{\vec {r}}^z S_{\vec {r}+\hat {y}}^z +S_{\vec {r}}^x
S_{\vec {r}+\hat {y}}^x> \ ,
\end{equation}
where angular brackets denote expectation value in the ground state of the
Hamiltonian in Eq.~(\ref{hamiltonian}). The paramagnetic term is given by
the equation 
\begin{equation}
\rho_{sy}^{para}=2E_{\theta} \ ,
\end{equation}
where $E_\theta$ is the coefficient of the $\theta^2$ term in the ground state
energy per site of the Hamiltonian in Eq.~(\ref{hamiltonian}) with a
perturbation 
\begin{equation}
\mathcal{H}^{para}=J_y\theta \sum_{\vec {r}} S^x_{\vec {r}} (S^z_{\vec {r}+%
\hat {y}} -S^z_{\vec {r}-\hat {y}}) \ .
\end{equation}
In order to calculate these quantities we introduce an Ising anisotropy\cite%
{oitmaa2006} by scaling all XY parts of the exchange interactions by a
factor $\lambda$. Then $\rho_{sx}$ and $\rho_{sy}$ can be calculated as a
power series in $\lambda$ for any value of the coupling anisotropy. Series
expansions for selected values of the anisotropy for the spin-half and
spin-one model are given in Table~\ref{TABLE 1} and Table~\ref{TABLE 2}
respectively.

The series are analyzed by Integrated Differential Approximants (IDA)\cite%
{oitmaa2006}. Before the analysis, a change of variable of the form $\sqrt{%
1-\lambda}=(1-y)$ has been introduced to remove leading singularities 
as $\lambda\to 1$. The results for the spin-half model are
shown in Fig.~\ref{fig:spinhalf} and the results for spin-one model are
shown in Fig.~\ref{fig:spinone}. In the 1D limit, the spin-stiffness
constant is know to be $0.25 J$ from exact calculations by Shastry and
Sutherland\cite{Shastry1990}. This value is clearly larger than the
square-lattice case where $\rho_s\approx 0.18 J$. Our results are more
accurate away from the 1D limit, but they clearly appear to approach the 1D
limit in a smooth and monotonic manner. For the spin-one case it is known
that the Ne\'{e}l order disappears at an anisotropy ratio of approximately $%
J_y/J_x=0.01$\cite{Affleck1989,Pardini2008}. The transition should be in the
universality class of the 3D Heisenberg model. The spin-stiffness should
vanish at the transition with an exponent of $\nu=0.7$\cite{Sachdev1995}.
The fit shows that the data agrees well with these expectations.

For the spin-half case we also fit the small anisotropy regime to a linear
behavior. The results are shown in Fig.~\ref{fig:alpha} where they are compared to the spin-wave results 
discussed in the next section. We find that the anisotropy can be expressed as 
\begin{equation}  \label{linearfit}
\rho_{sx}/\rho_{sy}=1+\kappa (J_x/J_y-1) \ ,
\end{equation}
where $\kappa=1.8$. This deviates significantly from the naive
expectation $\kappa=1$.

\section{Spin-wave calculation}\label{SW}

In this Section we consider the self-consistent version of spin-wave
theory\cite{SSWT,SSWT1,takahashi}. To apply this approach we subdivide the lattice
into sublattices $A$ and $B$ and use the Dyson-Maleev representation for
spin operators on each sublattice,%
\begin{eqnarray}
S_{i}^{+} &=&\sqrt{2S}a_{i}\,,\;S_{i}^{z}=S-a_{i}^{\dagger }a_{i},\;i\in A
\label{BKJa} \\
S_{i}^{-} &=&\sqrt{2S}(a_{i}^{\dagger }-\frac{1}{2S}a_{i}^{\dagger
}a_{i}^{\dagger }a_{i})  \notag
\end{eqnarray}%
and 
\begin{eqnarray}
S_{i}^{+} &=&\sqrt{2S}b_{i}^{\dagger }\,,\;S_{i}^{z}=-S+b_{i}^{\dagger
}b_{i}^{{}},\;i\in B  \label{BKJb} \\
S_{i}^{-} &=&\sqrt{2S}(b_{i}^{{}}-\frac{1}{2S}b_{i}^{\dagger
}b_{i}^{{}}b_{i}^{{}}),  \notag
\end{eqnarray}%
where $a_{i}^{\dagger },a_{i},$ and $b_{i}^{\dagger },b_{i}$ are the Bose
operators. Introducing the operators 
\begin{equation}
B_{i}=\left\{ 
\begin{array}{cc}
a_{i} & i\in A \\ 
b_{i}^{\dagger } & i\in B%
\end{array}%
\right.   \label{BC}
\end{equation}%
and decoupling the four-boson terms in the Hamiltonian into all possible
two-boson combinations, we derive (see Refs.~\onlinecite{SSWT,SSWT1})%
\begin{equation}
\mathcal{H}_{\text{SSWT}}=\sum_{i,\delta }J_{\delta }\gamma _{\delta }\left(
B_{i}^{\dagger }B_{i}-B_{i+\delta }^{\dagger }B_{i}\right)   \label{HSSWT} \ ,
\end{equation}%
where $\delta =x,y$ correspond to the nearest neighbour sites in the $\textit{x}$ and $\textit{y}$ 
directions, 
\begin{equation}
\gamma _{\delta }=\overline{S}+\langle a_{i}b_{i+\delta }\rangle 
\label{ksia}
\end{equation}%
are the short-range order parameters and
\begin{equation*}
\overline{S}=\langle S_{i\in A}^{z}\rangle =-\langle S_{i\in B}^{z}\rangle 
\end{equation*}%
is the sublattice magnetization. Diagonalizing the Hamiltonian (\ref{HSSWT})
one finds the self-consistent equations at $T=0$%
\begin{eqnarray}
\gamma _{\delta } &=&\overline{S}+\sum_{\mathbf{k}}\frac{\Gamma _{\mathbf{k}}%
}{2E_{\mathbf{k}}}\cos k_{\delta }  \notag \\
\overline{S} &=&S+1/2-\sum_{\mathbf{k}}\frac{\Gamma _{0}}{2E_{\mathbf{k}}} \ ,
\label{EqSSWT}
\end{eqnarray}%
where the antiferromagnetic spin-wave spectrum has the form 
\begin{equation}
E_{\mathbf{k}}=\sqrt{\Gamma _{\mathbf{0}}^{2}-\Gamma _{\mathbf{k}}^{2}} \ ,
\label{Ek}
\end{equation}%
with 
\begin{equation}
\Gamma _{\mathbf{k}}=2\left( J_{x}\gamma _{x}\cos k_{x}+J_{y}\gamma _{y}\cos
k_{y}\right) 
\end{equation}%
and $\Gamma _{\mathbf{0}}\equiv \Gamma _{\mathbf{k}=0}$ (we assume here that
the ground state is antiferromagnetically ordered, otherwise a bosonic chemical
potential $\mu \neq 0$ should be introduced in the dispersion (\ref%
{Ek}) to fulfill the condition $\overline{S}=0$, see Refs.~\onlinecite{SSWT,SSWT1}).
The parameters $\gamma _{\delta }$ are simply related to the spin
correlation function at the nearest-neighbor sites by
\begin{equation}
\gamma _{\delta }=|\langle \mathbf{S}_{i}\mathbf{S}_{i+\delta }\rangle
|^{1/2}.  \label{CFF}
\end{equation}%
The spin-wave stiffnes along the $\textit{x}$ and $\textit{y}$ axes is expressed
through these parameters as%
\begin{equation}
\rho _{s\delta }=J_{\delta }S\overline{S}\gamma _{\delta }  \label{rss}
\end{equation}%
The results obtained according to 
Eqs.~(\ref{EqSSWT}) and (\ref{rss}) are shown in Figs.~\ref{fig:spinhalf},~\ref{fig:spinone}. 
While for $S=1$ the spin-wave results are close to those obtained by series expansions over the
entire parameter range, for $S=1/2$ at large anisotropy, the two techniques give 
results for $\rho _{sy}$ which are qualitatively different. 
This difference is most probably due to topological
excitations within the half-integer spin chains, which are not considered by 
spin-wave theory. For $S=1$, at the transition to the Haldane phase,
we find the stiffness exponents to be $\nu_x=0.50$ and $\nu_y=1.07$, in
qualitative agreement with series expansions.

For $S=1/2$, a quantitative discrepancy between spin-wave theory and
series expansions is visible already at small anisotropies. While series expansion 
and spin-wave stiffness along the weaker exchange couplings axis ($J_y$) remain very close
to each other, a discrepancy arises from the stiffness along the stronger coupling direction. 
We find the coefficient in the linear fit (\ref{linearfit}) $\kappa _{\text{SW}}=1.3,$ somewhat lower
than what found by series expansion, as shown in Fig~\ref{fig:alpha}. It is
possible that part of the difference is due to numerical inaccuracies 
or high order effects in $1/S$. However, with
increasing anisotropy the difference is not just quantitative. It becomes
qualitative and it implies the onset of new physics for the
spin-half case associated with the Berry phase terms~\cite{Sachdev1995}.

\section{Pinning}\label{pinning}

We base our considerations on the theory of underdoped cuprates suggested in Ref.~\onlinecite{Milstein}.
According to this theory, the ground state of an
underdoped uniformly doped cuprate is a spin spiral, spontaneously directed
along the (1,0) or the (0,1) direction. At sufficiently small doping, $x < x_c$, the
spiral has a static component while at $x > x_c$ it is fully dynamic.
Here $x$ is the concentration of holes in CuO$_2$ plane. According to this
picture, the electronic liquid crystal observed in 
Ref.~\onlinecite{Hinkov2007,Hinkov2008} is the 
mostly dynamic spin spiral
which may still have some small static component. 
For Sr doped La$_{2}$CuO$_4$, the value of the
critical concentration is $x_c\approx 0.11$, and for YBa$_2$Cu$_3$O$%
_{6.+y}$ is $x_c\approx 0.09$.
The absolute value of the wave vector of the spin spiral (static or dynamic)
is given by
\begin{equation}  \label{Q}
Q=\frac{g}{\rho_s}x \ .
\end{equation}
Here $\rho_s \approx 0.18 J$ is the spin-stiffness of the initial Heisenberg
model~\cite{Singh1989}, $J\approx 130meV$ is the antiferromagnetic exchange parameter of the
model, and $g$ is the coupling constant for the interaction between mobile holes
and spin waves. We set the spacing of the tetragonal lattice equal to unity,
so the the wave vector $Q$ is dimensionless. To fit the neutron scattering
experimental data to the position of the incommensurate structure in Sr doped
single layer La$_{2}$CuO$_4$, we need to set $g\approx J$, and to fit similar
data for double layer YBa$_2$Cu$_3$O$_{6.+y}$, we need to set $g\approx 0.7J$%
. It is not clear yet why the values of $g$ for these compounds are slightly
different, but for purposes of the present work this difference is not
important. The coupling constant $g$ was calculated within the extended t-J
model \cite{Igarashi1992,Sushkov2004}. The result is $g=Zt$ where $t$ is the
nearest site hopping matrix element and $Z$ is the quasihole residue. It is
known that in cuprates $t\approx 3J$ and $Z\approx 0.3$. Thus the calculated
value of the coupling constant, $g\approx J$, agrees well with that found by
fitting of experimental data. The ground state energy of the spin spiral state
consists of two parts. The spin spiral with the wave vector $Q$ gives rise to
the gain $-gQ$ in the kinetic energy of a single hole. On the other hand the
spiral costs the spin elastic energy $\rho_sQ^2/2$. So the total balance is $%
E=\rho_sQ^2/2-xgQ$, and minimization with respect to $Q$ gives the wave
vector (\ref{Q}) and the energy per elementary cell
\begin{equation}  \label{E1}
E=\rho_sQ^2/2-xgQ=-\frac{Z^2t^2x^2}{2\rho_s} \ .
\end{equation}
There are also quantum corrections
to this energy, but they are small and hence not important for our purposes%
\cite{Milstein}. Note, that Eq.~(\ref{E1}) is valid for both $x
<x_c $ and $x >x_c$, assuming that $x$ is not large.

Up to now we have disregarded the anisotropy assuming a perfect square
lattice. To analyze anisotropy in the spiral direction we have to replace 
\begin{equation}
t\to t(1\pm\epsilon) \ ,
\end{equation}
where $\epsilon$ is due to the lattice deformation,
so $t_a=t(1+\epsilon)$ and $t_b=t(1-\epsilon)$.
The antiferromagnetic exchange, $J\propto t^2/U$,
also becomes anisotropic, $J_a=J(1+ 2\epsilon)$, $J_b=J(1- 2\epsilon)$.
Hence the spin-stiffness is replaced
by $\rho_s \to \rho_s(1\pm 2\kappa\epsilon)$, where $\kappa \approx 1.8$ has
been calculated above. 
Now we can see how the lattice deformation influences the spiral energy
(\ref{E1}). In the case when ${\bm Q}$ is directed along the $a^*=(1,0)$
we have to replace in (\ref{E1}) $t \to t_a$ and $\rho_s \to \rho_{sa}$;
and  in the case when ${\bm Q}$ is directed along the $b^*=(0,1)$
we have to replace in (\ref{E1}) $t \to t_b$ and $\rho_{s} \to \rho_{sb}$.
Note that the quasiparticle residue $Z$ is a scalar property and therefore
it is independent of direction of ${\bm Q}$.
Altogether, with account of the anisotropy, the energy (%
\ref{E1}) is replaced by 
\begin{equation}  \label{E2}
E\to-\frac{g^2x^2}{2\rho_s}\left[1\mp 2(\kappa-1)\epsilon\right] \ .
\end{equation}
The minus sign corresponds to $\mathbf{Q}$ directed along the 
$a^*=(1,0)$ axis, and the plus sign corresponds to $\mathbf{Q}$ directed 
alone the $ b^*=(0,1)$ axis. 
Interestingly, without the spin-quantum-fluctuations effect (i. e. if $%
\kappa=1$) the anisotropy in energy disappears.

Since $a^*< b^*$ it is most natural to assume that $t_a > t_b$, This means
that $\epsilon > 0$. This point is supported by the LDA calculation
performed in Ref.~\onlinecite{Andersen1995}. In this case, according to Eq. (\ref%
{E2}), the energy of the state with $\mathbf{Q}$ along the $a^*$-axis is higher
than that with $\mathbf{Q}$ along the $b^*$-axis. This disagrees with the 
experimental data in Ref.~\onlinecite{Hinkov2007,Hinkov2008}. However, the anisotropy of the
hopping matrix element $t$ is not straightforward. There are two competing
contributions to $\epsilon$. The first one is related to the lattice
deformation and is positive. The second one is related to oxygen chains
that are present in YBa$_2$Cu$_3$O$_{6.45}$ and is
negative. In principle it is possible for the negative contribution to win, making
 $\epsilon$ negative~\cite{PC}. The neutron scattering
anisotropy has been previously discussed within the Pomeranchuk instability
scenario~\cite{Yamase2006}. This is probably not
sufficient to explain the newest data, see discussion in Ref.~\onlinecite%
{Yamase2}. However, it is interesting to note that, to explain the sign of
the pinning, the Pomeranchuk scenario also requires a negative $\epsilon$.
Anyway, for further numerical estimates, we will assume
\begin{equation}
\epsilon \approx -0.02
\end{equation}
The absolute value is consistent with the 1\% lattice deformation and the
sign has been discussed above. In this case, according to Eq. (\ref{E2}),
the energy of the state with $\mathbf{Q}$ along the $a^*$-axis is lower and
this is consistent with experimental data. The direction of pinning energy at $%
x=0.09$ reads 
\begin{equation}
\epsilon(\kappa-1) \frac{g^2x^2}{\rho_s}\sim 5\times 10^{-2}meV
\end{equation}
This is the pinning energy per Cu site and it is a pretty strong pinning.
For comparison, the pinning energy of spin to the orthorhombic b-direction
in undoped La$_{2}$CuO$_4$ is just $\sim 1.5\times 10^{-3}meV$. Assuming that
the correlation length is at least comparable with the period of the spin
spiral, $\xi \sim 2\pi/Q \sim 17$, we find that the total pinning energy per
correlation unit is $\sim 5\times 10^{-2}\xi^2 \sim 15meV$, which is a significant energy scale.

\section{conclusions}\label{conclusions}
In this paper, we have studied the spin-stiffness constants for spatially
anisotropic spin-half and spin-one Heisenberg models using series expansions
and self-consistent spin-wave theory. The theoretical results have been of interest
in themselves and show the importance of Berry
phase interference terms in anisotropic square-lattice models.

Our primary motivation for the study has been to understand the phenomena
of electronic liquid crystal and its pinning in high temperature
superconductors. We find that quantum interference effects significantly enhance
the spin-stiffness anisotropy and this can provide the primary
mechanism for the pinning of the liquid crystal direction. We have provided a detailed quantitative 
account of the pinning energy in YBCO.

\acknowledgments
We are grateful to O.~K.~Andersen, V.~Hinkov, B.~Keimer, and
H.~Yamase for very important discussions and comments.

\end{document}